\documentstyle[12pt,texdraw]{article}
\def\lsim{\mathrel{\rlap{\raise 2.5pt \hbox{$<$}}\lower 2.5pt}}
\def\gsim{\mathrel{\rlap{\raise 2.5pt \hbox{$>$}}\lower 2.5pt}}

\begin{document}
\thispagestyle{empty}
\begin{flushright}
UNIL-TP-6/95, hep-ph/9510254\\
\end{flushright}
\begin{center}
{\bf{\Large Low Energy Sum Rules For Pion-Pion \\
Scattering and Threshold Parameters$^*$}}
\vskip 0.5cm
B. Ananthanarayan$^{**}$\\
D. Toublan\\
G. Wanders\\
Institut de physique th\'eorique, Universit\'e de Lausanne,\\ [-2mm]
CH 1015, Lausanne, Switzerland.\\
\vskip 1.cm
\end{center}
\begin{abstract}
We derive a set of sum rules for threshold parameters of pion-pion
scattering whose dispersion integrals are rapidly convergent and
are dominated by S- and P-waves absorptive parts.  Stringent constraints
on some threshold parameters are obtained.
\end{abstract}


\noindent{\underline{\hspace{11.6cm}}}\\
\begin{small}
$^{*}$Work Supported by the Swiss National Science Foundation\\
$^{**}$Address from October 1, 1995: Institut f\"ur Theoretische
Physik, Universit\"at Bern, CH 3012, Bern, Switzerland
\end{small}

\setcounter{equation}{0}
\section{Introduction}

\bigskip

Pion-pion scattering is a fundamental strong interaction process
that is particularly well suited for theoretical investigations.
The pion is the lightest hadron  and the principles
of axiomatic field theory
lead to a wealth of rigorous results,
some of which have a direct physical relevance~\cite{Martin1}.
These results are consequences of analyticity,
unitarity and crossing symmetry (isospin violation effects
are ignored).
On another front, chiral perturbation theory
 provides an extension of the current algebra
techniques and produces explicit representations for the
low energy pion-pion scattering amplitudes~\cite{Gasser1}.
These amplitudes exhibit the required general properties within
their domain of validity and their specific structure reflects
the fact that the pion is a Goldstone boson associated with the
spontaneous  breaking of the axial symmetry of the
massless quark limit of QCD~\cite{Leutwyler1}.

Unfortunately, experimental information on pion-pion scattering is
hard to obtain.  Phase shift analyses for the S- and P-waves are
available above threshold and up to 1400 MeV~\cite{Martin2}
but large uncertainties prevail for the threshold parameters
(scattering lengths and effective ranges)~\cite{Nagels}.
This is an awkward state of affairs since these parameters
play a central role in chiral perturbation theory.  The
situation can be improved by constructing solutions of
Roy's rigorous partial wave equations~\cite{Roy} which are
consistent with experimental information~\cite{Basdevant1}.
However this procedure does not fix the threshold parameters
uniquely.

In this paper we present an alternative approach to the problem
of low energy pion-pion scattering which does not resort to chiral
perturbation theory or to the Roy equations.  We derive constraints
on pion-pion threshold parameters which are consequences of exact
properties combined with the known low energy features of pion-pion
scattering.  Our tools are sum rules involving dispersion
integrals that are dominated by the low energy S- and P- waves.
The well known Olsson sum rules~\cite{Olsson} cannot be used since
they are sensitive to the high energy absorptive parts.
We have found three sum rules which fulfill our needs.  Their
dispersion integrals being dominated by low energy contributions,
depend significantly on threshold parameters and this dependence
cannot be ignored.  This leads us to the following strategy:
the S- and P- wave absorptive parts occuring in the integrands are
parametrized to reproduce the main characteristics of the low energy
cross-sections with the scattering lengths and effective ranges
as free parameters.  The parametrization we use has been proposed
by Schenk~\cite{Schenk}.  The sum rules become non-linear equations for
the S- and P- wave threshold parameters and a combination of
D- wave scattering lengths.  We show that the solutions of
these equations which are
compatible with the data are confined to a rather small portion
of the experimentally allowed domain.  This is our main result
and it establishes the relevance of our sum rules.  One may hope
that the expected improved data~\cite{Bijnens1} will allow a detailed
check of their implications.  Furthermore, the sum rules presented
here could be used as a tool to estimate corrections
to certain one loop predictions of chiral perturbation theory
{}~\cite{Anant1}.

We derive our sum rules in Section 2 using a crossing symmetric
decomposition of the definite isospin amplitudes into an
S- and P- wave term and a higher
waves contribution.  Their implications are established in
Section 3 by means of quadratic and linear fits of the
equations for the threshold parameters.  The
constraints are discussed and compared with chiral perturbation
theory results in Section 4.

\bigskip

\setcounter{equation}{0}
\section{Low energy Sum Rules}

\bigskip

We explain in this section how one obtains the approximate
relations between the threshold parameters and low energy
S- and P- wave absorptive parts which are at the basis of
our analysis.  We also present exact counterparts of these
relations which include the complete absorptive parts.

We exploit the quite remarkable fact established some time
ago~\cite{Basdevant1,Basdevant2},
that there is a set of analytic amplitudes
\raisebox{-2.5mm}{$\stackrel{\textstyle\hat{T}}{\sim}$} which have the exact
S- and P- wave absorptive parts, are crossing symmetric and respect
the Froissart bound.  These unique amplitudes are given by:

\begin{eqnarray}
& \displaystyle
\raisebox{-2.5mm}{$\stackrel{\textstyle\hat{T}}{\sim}$}(s,t,u) =
 {1\over 4} (s + t C_{st} + u C_{su})
\raisebox{-2.5mm}{$\stackrel{\textstyle a_0}{\sim}$}
+{1 \over \pi} \int_4^\infty {dx \over x(x-4)}  & \nonumber
\\
 \nonumber \\
& \displaystyle
\cdot \left\{ \left[ {s(s-4) \over  (x-s)} +{t(t-4)\over (x-t)}C_{st}
+{u(u-4)\over (x-u)}C_{su}\right] {\rm Im}
\raisebox{-2.5mm}{$\stackrel{\textstyle f_0}{\sim}$}(x) \right.
 & \\
\nonumber \\
& \displaystyle \left. +3\left[{s(t-u)\over x-s} + {t(s-u) \over x-t}C_{st}
+{u(t-s) \over x-u} C_{su} \right] {\rm Im}
\raisebox{-2.5mm}{$\stackrel{\textstyle f_1}{\sim}$}
(x)\right\} & \nonumber
\end{eqnarray}
Our notations are standard:
\begin{equation}
\raisebox{-2.5mm}{$\stackrel{\textstyle T}{\sim}$}=\left(\begin{array}{c}
T^0 \\ T^1 \\ T^2
\end{array}\right),
\raisebox{-2.5mm}{$\stackrel{\textstyle a_0}{\sim}$}=\left(\begin{array}{c}
a^0_0 \\ 0 \\ a^2_0
\end{array}\right),
\raisebox{-2.5mm}{$\stackrel{\textstyle f_0}{\sim}$}=\left(\begin{array}{c}
f^0_0 \\ 0 \\ f^2_0
\end{array}\right),
\raisebox{-2.5mm}{$\stackrel{\textstyle f_1}{\sim}$}=\left(\begin{array}{c}
0 \\ f_1^1 \\ 0
\end{array}\right)
\end{equation}

Here $T^I$ designates the isospin $I$
$s$-channel amplitude, $f^I_l$ is its $l$-th
partial wave
amplitude, $a_0^I$ is an S- wave scattering length and
$C_{st}$ and $C_{su}$ denote the crossing matrices.
Our normalization of $T^I$ is such that its $s$-channel
partial wave expansion is
\begin{equation}
T^I(s,t,4-s-t)=\sum_{l=0}^\infty (2l+1) f^I_l(s) P_l(1+{2t
\over s-4}),
\end{equation}
where $s$ is the square of the center of mass
energy and $t$ is the square of the momentum transfer,
both in units of $m_\pi^2$, $m_\pi$ being the pion mass.
Below the inelastic threshold the partial wave amplitudes
are given in terms of their phase shifts $\delta^I_l$:
\begin{equation}
f^I_l(s)=\sqrt{{s\over s-4}} e^{i \delta^I_l(s)} \sin
\delta^I_l(s),
\end{equation}

If we decompose the full amplitudes $T^I$ according to
\begin{equation}
\raisebox{-2.5mm}{$\stackrel{\textstyle T}{\sim}$}=
\raisebox{-2.5mm}{$\stackrel{\textstyle \hat{T}}{\sim}$}+
\raisebox{-2.5mm}{$\stackrel{\textstyle \overline{T}}{\sim}$}
\end{equation}
the absorptive parts of the second term
\raisebox{-2.5mm}{$\stackrel{\textstyle \overline{T}}{\sim}$}
turn out to be sums of partial waves absorptive parts with
$l\geq 2.$  It then follows that eq. (2.5) represents a
unique crossing symmetric decomposition of the amplitude
$T^I$ into an S- and P- wave contribution $\hat{T}^I$ and
a higher waves term $\overline{T}^I$.  We call $\hat{T}^I$
a truncated amplitude.  The properties of $\overline{T}^I$ are
consequences of rigorous twice subtracted fixed-t dispersion
relations.  Note that $\hat{T}^I$ is analytic in the three
variables $s,\ t$ and $u$ with cuts $[4,\infty)$.  Therefore
$\overline{T}^I$ is analytic too.  Clearly, neither
$\hat{T}^I$ nor $\overline{T}^I$ fulfills the unitarity
condition.  It should also be kept in mind that $\hat{T}^I$
does not carry the complete S- and P- waves; $\overline{T}^I$
contributes to the real parts of their amplitudes.

According to (2.5), every pion-pion threshold parameter is a
sum of a truncated part coming from \raisebox{-2.5mm}{$\stackrel
{\textstyle \hat{T}}{\sim}$} and a higher waves term due to
\raisebox{-2.5mm}{$\stackrel{\textstyle \overline{T}}{\sim}$}.
Definition (2.1) implies that the truncated S- wave scattering
lengths coincide with the full scattering lengths.  The
other truncated threshold parameters are obtained from (2.1)
as combinations of integrals over S- and P- wave absorptive
parts and S- wave scattering lengths.  We are looking for
threshold parameters, or linear combinations of such parameters,
which are well approximated by their truncated part.  That is
to say, we have to find combinations for which the higher waves
contribution is under control and can be assumed to be small.
We first try to do this for $\pi^0$-$\pi^0$ parameters.

Let $T(s,t,u)\equiv {1\over 3}(T^0(s,t,u)+2 T^2(s,t,u))$ be the
full $\pi^0$-$\pi^0$ amplitude.  According to (2.1) its truncated
version is:
\begin{eqnarray}
& \displaystyle \hat{T}(s,t,u)=a + {1 \over \pi}
\int_4^{\infty}  {dx \over x(x-4)} & \nonumber\\
 \\
& \displaystyle
\cdot \left[ {s(s-4)\over x-s}+{t(t-4)\over x-t}+
{u(u-4)\over x-u}\right] {\rm Im} f(x) & \nonumber
\end{eqnarray}
with $f={1\over 3}(f_0^0+2f_0^2)$, $a={1\over 3}(a_0^0+2a_0^2)$.
Eq. (2.6) gives for $t=0$:
\begin{eqnarray}
& \displaystyle {{\rm Re}\left(\hat{T}(s,0,4-s)-\hat{T}(4,0,0)
\right) \over s-4}=
& \nonumber \\
 \\
& \displaystyle
{s\over \pi} {\rm P}\int_4^\infty dx {1\over x(x-4)}
{2x-4 \over x(x-4)-s(s-4)}{\rm Im} f(x) & \nonumber
\end{eqnarray}

Threshold parameters specify the behaviour of a scattering
amplitude as $s\to 4$ from above.  Some care is required
in taking the limit of the integral in (2.7) since it
appears to diverge at first sight.  We have to exploit the
threshold behaviour of ${\rm Im} f$ which allows us to write:
\begin{equation}
{\rm Im} f(x)={1\over 16 \pi} \sqrt{x(x-4)} \sigma(x)
\end{equation}
with $\sigma$ regular at $x=4$ ($\sigma$ is the S-wave
$\pi^0$-$\pi^0$ total cross-section).  After insertion of
(2.8) into (2.7) one finds $\sigma(x)$ can be replaced
by $\left( \sigma(x)-\sigma(4) \right)$ in the integrand
if $s>4$, without changing the value of the integral.
This is due to the identity:
\begin{equation}
{\rm P}\int_4^{\infty}dv' {1\over \sqrt{v'-4}(v'-v)}=0,
\end{equation}
which is true if $v>4$.  The limit $s\to 4+$ can be taken
safely once this subtraction has been performed.

The limit of the left-hand side of (2.7) is equal to
$\hat{b}/4$, $\hat{b}$ being the truncated $\pi^0$-$\pi^0$
S- wave effective range.
We use the standard definition of scattering lengths
$a^I_l$ and effective ranges $b_l^I$:
\begin{equation}
{\rm Re} f_l^I(\nu)=\nu^l(a_l^I+b_l^I\cdot \nu+...)
\end{equation}
where $\nu$ is the square of the center of mass momentum,
$\nu\equiv(s-4)/  4$.  Using $\nu$ as the integration variable,
we obtain the following sum rule
\begin{equation}
\hat{b}={1\over 4\pi^2}\int_0^\infty d\nu {2\nu+1 \over (\nu(\nu+1))^{3/2}}
(\sigma(\nu)-\sigma(0))
\end{equation}

A second sum rule is obtained by combining the derivative of (2.6)
with respect to $t$ at threshold with (2.11).  It gives the
truncated D-wave scattering length $\hat{a}_{(2)}$ ($
a_{(2)}={1\over 3}
(a^0_2+2a_2^2)$):
\begin{equation}
\hat{a}_{(2)}
={1\over 60\pi^2}\int_0^\infty d\nu {\nu^{1/2} \over (\nu+1)^{5/2}}
\sigma(\nu)
\end{equation}

An important point is that we have sum rules not only for the truncated
$\hat{a}_{(2)}$ and $\hat{b}$ but also for the complete D- wave scattering
length $a_{(2)}$ and S-wave effective range $b$~\cite{Wanders1}.
The latter are consequences of the exact analyticity properties of the full
amplitudes $T^I$ and crossing symmetry.  Combining them with (2.11)
and (2.12) one obtains the decompositions:
\newpage

\begin{eqnarray}
& \displaystyle b=\hat{b}+\overline{b}, & \nonumber  \\
&  \displaystyle a_{(2)}=\hat{a}_{(2)}+\overline{a}_{(2)} &
\end{eqnarray}
the higher wave contributions given by
\begin{eqnarray}
& \displaystyle
\overline{b}=
{1\over 4\pi^2}\int_0^\infty d\nu {2\nu+1 \over (\nu(\nu+1))^{3/2}}
(\overline{\sigma}(\nu)), & \nonumber \\
\\
& \displaystyle
\overline{a}_{(2)}={1\over 60\pi^2}\int_0^\infty d\nu\left[
{\nu^{1/2} \over (\nu+1)^{5/2}}
\overline{\sigma}(\nu)+8\pi{2\nu+1\over (\nu(\nu+1))^2}
{\partial\over \partial t} \overline{A}(\nu,0)\right], & \nonumber
\end{eqnarray}
where $\overline{A}(\nu,t)$ is the absorptive part of $\overline{T}$
and $\overline{\sigma}(\nu)$ is the higher waves contribution to
the total cross section: $\overline{\sigma}(\nu)=4\pi(\nu(\nu+1))^{
-1/2} \overline{A}(\nu,0)$.

The weight functions appearing in (2.11), (2.12) and (2.14) favor the low
energy parts of the integrals.  As the higher partial waves are small at
low energies we may expect that $\overline{a}_{(2)}$
and $\overline{b}$
are small with respect to $\hat{a}_{(2)}$
and $\hat{b}$.  Indeed, one finds
that the contribution of the D- wave resonance $f_2(1270)$~\cite{pdg}
to $\overline{a}_{(2)}$ and $\overline{b}$ is of the order of $10\%$
of the accepted values of $a_{(2)}$ and $b$~\cite{Nagels}.
This indicates that $\overline{a}_{(2)}$ and $\overline{b}$ are in fact
small compared to $\hat{a}_{(2)}$ and $\hat{b}$ but not negligibly small.
Therefore an evaluation of $\hat{b}$ and $\hat{a}_{(2)}$ gives only a
relatively crude estimate of the full parameters $b$ and $a_{(2)}$.
As it is our ambition to derive more precise predictions, we have to
find sum rules giving low energy parameters which are well approximated
by truncated integrals.  We must admit that our $\pi^0$-$\pi^0$
sum rules do not really meet our requirements.

In order to achieve our aims, we have to work with amplitudes having
the same analyticity properties as the scattering amplitudes $T^I$
but a better asymptotic behaviour.  The $t-u$ antisymmetry of
$T^1(s,t,u)$ implies that
\begin{equation}
H(s,t,u)={T^1(s,t,u)\over t-u}
\end{equation}
is such an amplitude.
Furthermore, we find that the following three functions
are suitable for our purposes:
\begin{eqnarray}
& \displaystyle F_{\stackrel{1}{2}}(s,t)=H(t,u,s)\mp
\left( H(s,t,u)+H(u,s,t) \right ) & \nonumber \\
\\
& \displaystyle F_3(s,t) = H(t,u,s) + H(u,s,t) & \nonumber
\end{eqnarray}

Proceeding as in the $\pi^0$-$\pi^0$ case, one finds, for the
truncated versions of $\hat{F}_\alpha$ of the $F_\alpha$'s:
\begin{eqnarray}
& \displaystyle -96\lim_{s \to 4+} {\partial\over \partial s}
\hat{F}_1(s,0)= & \nonumber \\
& \displaystyle \left (2\hat{a}_0^0-5\hat{a}^2_0\right) -18 \hat{a}^1_1
+18\hat{b}^1_1= & \\
& \displaystyle {1\over 4 \pi^2}\int_0^\infty
d\nu \left[{1\over \nu^{1/2} (\nu+1)^{5/2}}\left(2\sigma_0(\nu)
-5\sigma_2(\nu)\right)+ \right. & \nonumber \\
& \displaystyle \left.
3 {\nu^2+4\nu+2 \over (\nu(\nu+1))^{5/2}}\sigma_1(\nu)
\right] & \nonumber
\end{eqnarray}
\begin{eqnarray}
& \displaystyle -96\lim_{s \to 4+} {\partial\over \partial s}
\hat{F}_2(s,0)= & \nonumber \\
& \displaystyle  3\left( 2\hat{a}_0^0-5\hat{a}^2_0\right) -2
\left( 2\hat{b}_0^0-5\hat{b}^2_0\right)
-18\hat{b}^1_1= & \\
& \displaystyle {1\over 4 \pi^2}\int_0^\infty
d\nu {1\over (\nu(\nu+1))^{3/2}} \left[ -{3\nu+2\over \nu+1}
\left(2\sigma_0(\nu)
-5\sigma_2(\nu)\right) \right. & \nonumber \\
& \displaystyle \left.
+2(\nu+1)\left(2\sigma_0(0)
-5\sigma_2(0)\right)-3{3\nu^2+6\nu+2 \over \nu(\nu+1)}\sigma_1(\nu)
\right ] & \nonumber
\end{eqnarray}
\begin{eqnarray}
& \displaystyle -48\lim_{s \to 4+} \left({1\over \nu}{\partial \over \partial
t}\hat{F}_3(s,0)\right)  =& \nonumber \\
& \displaystyle
\left( 2\hat{a}_0^0-5\hat{a}^2_0\right)-18 \hat{a}^1_1
+30\left( 2\hat{a}_2^0-5\hat{a}^2_2\right)= &  \\
& \displaystyle {1\over 4\pi^2} \int_0^\infty d\nu
{\nu \over (\nu(\nu+1))^{5/2}} \left [ \nu
\left( 2\sigma_0(\nu) -5\sigma_2(\nu) \right) +3 \sigma_1
(\nu) (\nu-2) \right ] & \nonumber
\end{eqnarray}

As in (2.7), $\sigma_I(\nu)=4\pi (2l+1)(\nu(\nu+1))^{1/2}{\rm Im}
f^I_l(\nu)$, $l=0$ for $I=0,\ 2$ and $l=1$ for $I=1$.

One finds that
\[
\lim_{s\to 4+}{1\over \nu}{\partial \over \partial t}
H(s,0)
\]
gives a sum rule of the same type for $a^1_3$, the
$I=1$, F- wave scattering
length, which we shall not use.

Again there are exact sum rules for the higher waves contributions to
the combinations of threshold parameters appearing in (2.17)-(2.19).
Such sum rules can be obtained by a straightforward application of the
technique used for the $\pi^0$-$\pi^0$ amplitude in Ref.~\cite{Wanders1}
to $F_1$ and another totally symmetric amplitude constructed in
Ref.~\cite{Roskies}.  We display the results without going
through the proofs:

\begin{eqnarray}
& \displaystyle -18 \overline{a}^1_1+18 \overline{b}^1_1=
& \nonumber \\
& \displaystyle {1\over 4 \pi^2}\int_0^\infty
d\nu \left[{1\over \nu^{1/2} (\nu+1)^{5/2}}\left(2\overline{\sigma}
^0(\nu)
-5\overline{\sigma}^2(\nu)\right)+ \right. &  \\
& \displaystyle \left.
 3{\nu^2+4\nu+2 \over (\nu(\nu+1))^{5/2}}\overline{\sigma}^1(\nu)
\right] & \nonumber
\end{eqnarray}
\newpage

\begin{eqnarray}
& \displaystyle -2\left(\overline{b}^0_0-\overline{b}^2_0
\right) -18 \overline{b}^1_1= & \nonumber \\
& \displaystyle {1\over 4 \pi^2}\int_0^\infty
d\nu {1\over (\nu(\nu+1))^{3/2}} \left[ -{3\nu+2\over \nu+1}
\left(2\overline{\sigma}^0(\nu)
-5\overline{\sigma}^2(\nu)\right) \right. &  \\
& \displaystyle \left.
-3
{3\nu^2+6\nu+2 \over \nu(\nu+1)}\overline{\sigma}_1(\nu)
\right ] & \nonumber
\end{eqnarray}
\begin{eqnarray}
& -18 \overline{a}^1_1 + 30 \left ( 2\overline{a}^0_2 -5\overline{a}^2_2
\right) = & \nonumber \\
& \displaystyle {1\over 4\pi^2} \int_0^\infty
d\nu \left[{1\over \nu^{1/2}(\nu+1)^{5/2}}\left(
2\overline{\sigma}^0-5\overline{\sigma}^2+3\overline{\sigma}^1
\right)-{6(2\nu+1)\over (\nu(\nu+1))^{5/2}}\overline
{\sigma}^1 \right. &  \\
& \displaystyle \left. +{16\pi \over (\nu(\nu+1))^2}
{\partial \over \partial t}\left(2\overline{A}^0-5\overline{A}^2
+3\overline{A}^1\right) \right] & \nonumber
\end{eqnarray}
In these integrals, $\overline{\sigma}^I$ and $\overline{A}^I$ are
the higher waves contributions to the isospin $I$ total cross-section
and absorptive part.  Furthermore, the fact that $\overline{a}^0_0=
\overline{a}^2_0=0$ has been taken into account.

We now have weight functions decreasing more rapidly than the
corresponding $\pi^0$-$\pi^0$ weight functions we had before.
This results in the contributions of $f_2(1270)$ to the integrals
reduced to the order of $1\%$ of the expected values of the
combinations of complete threshold parameters.  As a consequence,
we can now transform eqs. (2.17)-(2.19) into reliable approximate
sum rules by replacing the truncated parameters in the left-hand
sides by their complete counterparts.  In this manner, we arrive
at a very helpful tool for the analysis of low energy pion-pion
scattering, which will be established in the next Section.

It is worth noting that the existence of the truncated amplitudes
defined in eq. (2.1) is due to the convergence of twice subtracted
dispersion relations.  Their uniqueness implies that crossing symmetry
alone does not constrain the S- and P- wave absorptive parts.
Furthermore, the uniqueness of $\hat{T}^I$ implies the uniqueness
of the right hand sides of (2.17)-(2.19).  This does not apply to
the right hand sides of (2.20)-(2.22).  As a matter of fact, there
are various inequivalent methods leading to different
expressions for the right hand sides.
The fact that the  values of these expressions
have to be equal, leads to constraints on the higher
waves absorptive parts, in contrast to those of the S- and P- waves.

\setcounter{equation}{0}
\section{Transforming the sum rules into equations for
threshold parameters}

\bigskip

The presently available data do not allow a reliable evaluation of
the integrals appearing in our sum rules.  In particular they are
quite sensitive to the values of the threshold parameters.
If these quantitites were to be known precisely, one could use the
sum rules to test their consistency with field theoretic
predictions.
In the present situation some threshold parameters are only
poorly known and the best one can possibly do is
to turn the sum rules into
non-linear equations for these parameters and determine if and
how their possible values are constrained.  To achieve this aim
in a simple way we require an analytic parametrization of the
S- and P- wave phase shifts, containing the
scattering lengths and effective ranges as
free parameters, and
reproducing their main known features above threshold and
below the $K-\overline{K}$ threshold.  A parametrization
has been provided by Schenk~\cite{Schenk} along
these lines, which we use
with the $I=1$ P- wave modified slightly in such a way that
it depends only on $a_1^1$ and $b_1^1$.
The explicit form of the parametrization
we shall use is:
\begin{eqnarray}
& \displaystyle \tan \delta_0^I(\nu)= & \nonumber \\
& \displaystyle
{\nu^{1/2}\over (\nu+1)^{1/2}}\left\{
a_I+[b_I-a_I/\nu_0^I+(a_I)^3]\nu \right\}{\nu_0^I \over
(\nu_0^I-\nu)},\  I=0, 2, & \\
& \displaystyle \tan \delta_1^1(\nu)=
{\nu^{3/2}\over(\nu+1)^{1/2}}\left\{
a_1+[b_1-a_1/\nu_\rho]\nu\right\}{\nu_\rho \over
(\nu_\rho-\nu)}  &
\end{eqnarray}
The S- and P- wave parameters have been relabelled:
$a_I=a^I_l,\ b_I=b_l^I,\ l=0$ if $I=0, 2$ and $l=1$ for $I=1$.
We take $\nu_0^0=8.5, \ \nu^2_0=-5.0$ as in Ref.~\cite{Schenk}
and $\nu_\rho=6.6$, which is the position of the $\rho(770)$
resonance.  Note that these representations for the
phase shifts ensure normal threshold behaviour.

Another representation of the S- and P- wave phase shifts may
be obtained from numerical solutions to the Roy equations
that are consistent with experimental data~\cite{Basdevant1}.
Nevertheless, the difference between this and the representation
we use has been found to yield a difference at the level of
a few percent in  the present analysis when the parameters in
(3.1) and (3.2) are correctly adjusted~\cite{Anant2}.

Once the cross-sections $\sigma_I$ determined by (3.1) and (3.2)
are inserted into the integrals of (2.17)-(2.19), these integrals
become non-linear functions of the 6 parameters $a_I$ and $b_I$.
When we evaluate these integrals numerically, we cut them
off at $\nu=11$ corresponding to a total energy of $970$ MeV,
contributions from higher energies being negligible.

We shall explore a restricted domain of the space spanned by
these parameters:
\begin{eqnarray}
& \displaystyle a_0\in (0,1),\ a_2\in (-0.1,0), \ a_1\in (0,0.1)
& \nonumber \\
& \displaystyle b_0\in(0,1),\ b_2 \in (-0.2,0),\ b_1 \in (0,0.02) &
\end{eqnarray}

The experimental data
for the 5 first parameters give~\cite{Nagels}:
\begin{eqnarray}
& \displaystyle a_0\in (0.21,0.31),\ a_2\in (-0.040,-0.016),
\ a_1\in (0.036,0.040)
& \nonumber \\
& \displaystyle b_0\in(0.22,0.28),\ b_2 \in
(-0.090,-0.074), &
\end{eqnarray}
while no experimental information on $b_1$ is available.
Note that these values are well inside the domain defined by (3.3).

Since the sum rules integrals are smooth functions of the parameters,
we approximate them in the domain defined by eq. (3.3) by least
square quadratic fits.  The fits $I_1,\ I_2$ and $I_3$ of the
integrals in (2.17), (2.18) and (2.19) have the form
($\alpha=1,2,3)$:
\begin{equation}
I_\alpha=C_\alpha+\sum_{I=0}^2 [R_{\alpha I} a_I + S_{\alpha I}
b_I + T_{\alpha I} (a_I)^2 + U_{\alpha I} (b_I)^2 + V_{\alpha I}
a_I b_I]
\end{equation}

The values of the coefficients are given in Table 1.  The
relative standard deviation of the fit (3.5) is less than
$4\%$ for all the integrals and the correlation
coefficients squared are all larger than $0.99965$.

With (3.5) the sum rules produce 3 equations of second degree
in the S- and P- wave parameters $a_I$ and $b_I$ and the D- wave
parameter
\begin{equation} \nonumber
A_2\equiv 2a^0_2-5a^2_2
\end{equation}
One may ask how many solutions of these equations are located
in the domain (3.3).  In order to find the answer, we observe
that the following sum rule
\begin{equation}
b_1-{5\over 3} A_2 ={1 \over 12\pi^2} \int_0^\infty
{3\nu+1 \over (\nu(\nu+1))^{5/2}}\sigma_1(\nu)
\end{equation}
is a consequence of (2.17)-(2.19).  Remarkably, its
integral depends only on the $I=1$ P- wave cross section.
The quadratic version of the integral is $(I_1-I_3)/18$.  One sees
that it depends only on the $I=1$ P- wave parameters;
no fictitious S- wave dependence is introduced through our
fit of the $I_\alpha$'s.  Solving the quadratic
approximation of (3.7) with respect to $b_1$, we find
\begin{equation}
b_1=-1.80+1.26 a_1 \pm [3.24-4.51 a_1 +3.72 (a_1)^2+6.31 A_2]^{1/2}
\end{equation}

Fits to the experimental data give~\cite{Nagels}
\begin{equation}
A_2=2 a_2^0-5 a^2_2=(0.275\pm 0.210) \cdot 10^{-2}
\end{equation}

If we constrain $A_2$ to this range,
we find that only the $+$ solution
in (3.7) belongs to the domain (3.3).
Introducing this solution into (2.18) and (2.19)
one gets two equations for $a_0, \ b_0, \ a_2,\ b_2$
and $a_1$ at fixed $A_2$.  It turns out that they
have only one solution compatible with (3.3) if $A_2$
is in the interval (3.9).  That is to say that if one
chooses any triplet among the first 5 of these parameters
which fulfills (3.3) and if $A_2$ is fixed according to
(3.9), then $b_1$ is determined by (3.8) and only one
value of the remaining pair of parameters obeys (3.3)
and is allowed by the sum rules (2.18) and (2.19).

This results leads naturally to the question of the
existence of sets of parameters which are compatible with
the sum rules as well as with the experimental data.
To answer this question, we restrict the S- and P- wave
parameters to the domain (3.4) and the physically unknown
$b_1$ to the interval $(0.002,0.010)$ by using
(3.8).  The sum rule integrals
behave nearly linearly in this domain and we simplify
our discussion by replacing the quadratic fits (3.5) by
linear ones.  The new correlation coefficients are larger
than 0.996.

The linearized version of (3.7) is
\begin{equation}
b_1=1.795 A_2 + 0.053 a_1 -0.001
\end{equation}
Using (3.4) and (3.8) this gives
\begin{equation}
b_1=0.006\pm 0.004.
\end{equation}
The $+$ solution in (3.7) gives the
same value, the large error coming mainly from the large
uncertainty on $A_2$ in (3.9).  As far as we know, we have here
the first determination of the $I=1$ P- wave effective range
based on sum rules.

In addition to (3.7) we have the 2 independent sum rules
(2.18) and (2.19).  Eliminating $b_1$ by means of (3.10) in their
linearized versions one is left with 2 linear equations relating
6 parameters.  We find it convenient to express these
equations in terms of $a_0,\ a_1,\ b_0, A_0,\ B_0$ and
$A_2$ where $A_0$ and $B_0$ are corrected versions of the differences
$(0.4 a_0 - a_2)$ and $(0.4 b_0 - b_2)$ appearing in the
left hand sides of the sum rules:
\begin{equation}
A_0\equiv 0.27 a_0 - a_2, \ B_0 \equiv 0.27 b_0 -b_2
\end{equation}
The linearized sum rules (2.18) and (2.19) can now be written as:
\begin{eqnarray}
& \displaystyle A_0-0.529 B_0=-0.001 a_0 +0.004 b_0 +0.797 a_1 -0.009
& \nonumber \\
& \displaystyle A_0 -0.689 B_0 -1.692 A_2=0.010 a_0 -0.002 b_0 -0.009
&
\end{eqnarray}

The equations above and eq. (3.10) express the constraints
imposed by our sum rules in a domain of threshold parameters consistent
with the data.  These constraints are analyzed in the next section.

\bigskip

\setcounter{equation}{0}
\section{Discussion of the constraints on threshold parameters}

\bigskip

It is con\-venient to dis\-cuss the impli\-cations of eq. (3.13)
in the $(A_0, B_0, A_2)$ space.  The experimental data
(3.4) define a domain $\Delta$ for these parameters:

\begin{eqnarray}
& \displaystyle
\Delta \equiv \left\{A_0\in(0.072,0.124),\ B_0 \in(0.134,0.166),\right.&
 \nonumber \\
& \displaystyle
\left. A_2\in (0.065\cdot 10^{-2}, 0.485\cdot 10^{-2})\right \} &
\end{eqnarray}
For given values of $a_0, b_0$ and $a_1$, the equations (3.13)
constrain the point with co-ordinates
 $(A_0,\ B_0,\ A_2)$ to a straight line
$d(a_0,b_0,a_1)$.  One finds that this line intersects $\Delta$ for
all values of $a_0,\ b_0$ and $a_1$ allowed by (3.4)
along a segment $\overline{d}(a_0, b_0, a_1)$ as shown in
Fig. 1.  The values of $(A_0,\ B_0,\ A_2)$ which are
compatible with experimental data and the sum rules define a
domain $\overline{\Delta}$ which is the union of the segments
 $\overline{d}(a_0, b_0, a_1)$ corresponding to all values of
$(a_0, b_0, a_1)$ in the domain (3.4).
As a consequence of our linearization, $\overline{\Delta}$
is a convex domain; it is shown
in Fig. 2; apart from eq. (3.10) it displays all the
information we have derived from our approximate sum rules.
The constraints we obtain are quite spectacular.  $\overline{\Delta}$
is a very narrow
prism bounded by four planes and truncated by the
faces of $\Delta$.  The faces $A_0=0.072$ and $A_0=0.124$ are
completely excluded and the same is nearly true for the
faces $A_2=0$ and $A_2=0.005$.  Whereas $B_0$ is unconstrained
there are strong correlations on the possible values of $A_0$
and $A_2$ at given $B_0$.  Fig. 3 shows the projections of
$\overline{\Delta}$ onto the $(A_0, A_2)$-, $(A_2,B_0)$- and
$(A_0,B_0)$- planes.  A very narrow strip is selected in the
$(A_0,B_0)$-plane.  The central experimental values are
slightly outside this strip.

If we restrict ourselves to the S- wave scattering lengths, we
see that $A_0$ is confined to the interval $(0.090,0.112)$.  This
defines a band bounded by
$|0.27 a_0 - a_2 -0.101|<0.011$  in the $(a_0, a_2)$ plane.
This band is shown in Fig. 4.  A similar band shows up in
many other analyses of pion-pion scattering;  the one used in
Ref.\cite{Nagels} is also shown in Fig. 4 together with the
rectangle compatible with the data.  Clearly, most of the
correlations encoded in the shape of $\overline{\Delta}$
are washed out by the projection onto this  $(a_0, a_2)$
 plane.  Despite this, the sum rules still impose
efficient constraints.

Finally we compare our results with the predictions of two
versions of chiral perturbation theory (CHPT), the so called
standard one (SCHPT)~\cite{Gasser1,Gasser2,Bijnens2}
and the generalized one (GCHPT)~\cite{Stern1,Knecht1,Knecht2}.
Table 2 gives three sets of central values of threshold
parameters obtained from various one loop pion-pion scattering
amplitudes~\cite{Bijnens2,Knecht1}.
They define points $P$ in the $(A_0,B_0,A_2)$-space whose locations
are indicated in Figs. 5a-c together with the relevant portions
of our allowed domain $\overline{\Delta}$.  The SCHPT point
$P_{\displaystyle a}$
is just at the border of this domain although its
$a_0$ value is slightly below the interval allowed in
(3.4).  The GCHPT points $P_{\displaystyle b}$
and $P_{\displaystyle c}$ are outside
$\overline{\Delta}$ whereas their $(a_0,b_0,a_1)$ obey (3.4).  In
other words, one loop SCHPT threshold parameters can be considered
consistent with our sum rules combined with Schenk's parametrization
while this is not quite true for GCHPT.

When discussing the positions of the points $P$ with respect
to sections of $\overline{\Delta}$ we do not take into account the
CHPT values of $a_0,\ b_0$ and $a_1$.  We do this in a second
exercise: inserting the values of $a_0,\ b_0,\ a_1$ and $B_0$
from Table 2 into the constraints (3.13) we obtain values of
$A_0$ and $A_2$ also given in Table 2.
They define new points $Q$ in Figs. 5a-c.
That is to say, these points are produced by our
sum rules implemented with Schenk's parametrization (3.1)-(3.2)
taken at CHPT values of $a_0,\ b_0$ and $a_1$.  They are inside their
sections of $\overline{\Delta}$, indeed as they must be, but do not
coincide with the points $P$ defined earlier.  It may be
inferred from the peculiar fact that one loop chiral amplitudes
fulfill our sum rules identically~\cite{Anant1} that the
discrepancy between the $Q$'s and $P$'s is essentially due to the
difference between Schenk's and the one loop chiral absorptive parts.
The former being certainly closer to the true ones above threshold
and at intermediate energies, we conclude that our results establish
the necessity of non-negligible higher order corrections to
one loop calculations.

Two loop computations in the framework of GCHPT are
presented in Ref.~\cite{Knecht2} and a sample of two loop
threshold parameters is displayed in Table 2.  This sample
defines two points $Q$ and $P$ which are practically identical
(Fig. 5d).  This spectacular improvement must come from the
two loop corrections to the absorptive parts and a larger
flexibility in the choice of effective coupling constants.
The circumstance that this choice is partly based on sum
rules may also be playing a role.  These are sum rules based
on twice subtracted dispersion relations involving high energy
contributions in contrast with the low energy sum rules analyzed
here.  For a check of sum rule (3.7), the values of $b_1$ as
obtained via eq. (3.10) from the CHPT data for $A_2$ and $a_1$
are given in Table 2.  Whereas the sum rule predictions differ
from CHPT values at one loop, the agreement is again excellent
at two loop GCHPT.

Our discussion of CHPT pion-pion scattering illustrates the
relevance of low energy sum rules.  They reveal definitely the
need of two loop corrections.  However, as no two loop results
obtained in the strict SCHPT framework are available at present,
we cannot tell whether our tools allow a discrimination between
that scheme and GCHPT.

Although our analysis is based on exact sum rules we have
had to make two major approximations which are not under precise
quantitative control.  First, the contributions from the
higher partial waves due to
\raisebox{-2.5mm}{$\stackrel{\textstyle\overline{T}}{\sim}$}
 in the decomposition
(2.3) have been neglected.
Second, we have played our game
using the very simple
analytic parametrization (3.1)-(3.2) for the S- and P-
waves.  An improved parametrization will modify the shape
of $\overline{\Delta}$ whereas an evaluation of the size of the
\raisebox{-2.5mm}{$\stackrel{\textstyle\overline{T}}{\sim}$}
 contributions would allow an estimation of the
uncertainties
coming from these contributions.  This would enlarge
 $\overline{\Delta}$.  Since our domain  $\overline{\Delta}$
is well inside $\Delta$, we believe that our results are
robust and will survive these improvements.

\bigskip

\noindent{\bf Acknowledgements:} We thank J. Gasser and
H. Leutwyler for their continuing interest in the subject and
discussions.  We thank M. Knecht for useful correspondence.

\newpage

\noindent{\bf Table Captions}

\bigskip

\noindent {\bf Table 1} Coefficients of the quadratic fits
(3.5) of the sum rules integrals in eqs. (2.17)-(2.19).

\bigskip

\noindent{\bf Table 2} Threshold parameters of chiral perturbation
theory (CHPT).  For each set of parameters the first column gives
central CHPT values and the second column gives sum rules predictions
based on the values found in the first column.  The values of
$A_0$ and $A_2$  in a first column define $P$
in Fig. 5 whereas those in the corresponding second column define $Q$.
{\bf (a)} Threshold parameters of a standard one loop amplitude carrying
the coupling constants $L_1, \ L_2$ and
$L_3$ given in the 4th column of Table 2 in
Ref.~\cite{Bijnens2},
{\bf (b)-(c)} threshold parameters of two generalized
one loop amplitudes displayed in Table 1 of~\cite{Knecht1} corresponding
to two values of the coupling constant $L_3$,
{\bf (d)} threshold parameters
of an extended two loop amplitude: 5th line in Table 1 of~\cite{Knecht2}.

\newpage

\noindent{\bf Figure Captions}

\bigskip

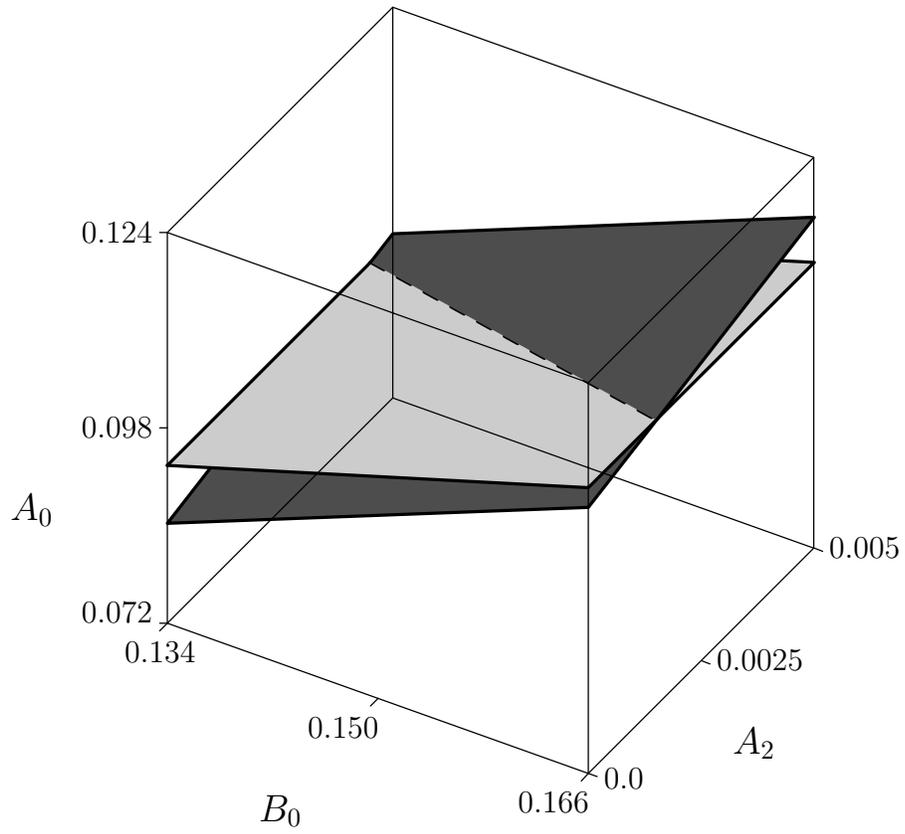
\noindent {\bf Fig. 1.}  The two planes in the
$(A_0,B_0,A_2)$- space defined by eq. (3.13) for the
central values of the right hand sides according to (3.4).
The points belonging to the intersection of these planes
which are inside the domain $\Delta$ [defined in eq. (4.1)]
are physically admissible, compatible with the sum rules and
the experimental central values of $(a^0_0, b^0_0, a^1_1)$.


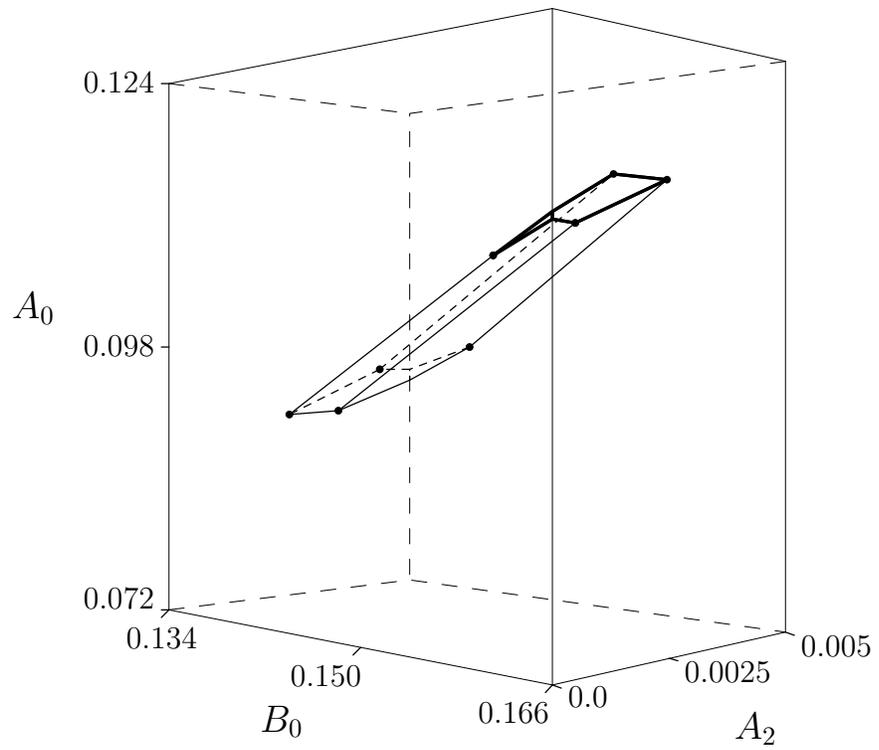
\noindent {\bf Fig. 2.}  The domain $\overline{\Delta}$ in the
$(A_0,B_0,A_2)$- space which is compatible with experimentally
allowed values of $(a^0_0, b^0_0, a^1_1)$
as defined in (3.4).  The vertices of $\overline{\Delta}$
are marked by dots:  they are at the intersection of a prism
with the faces of $\Delta$.


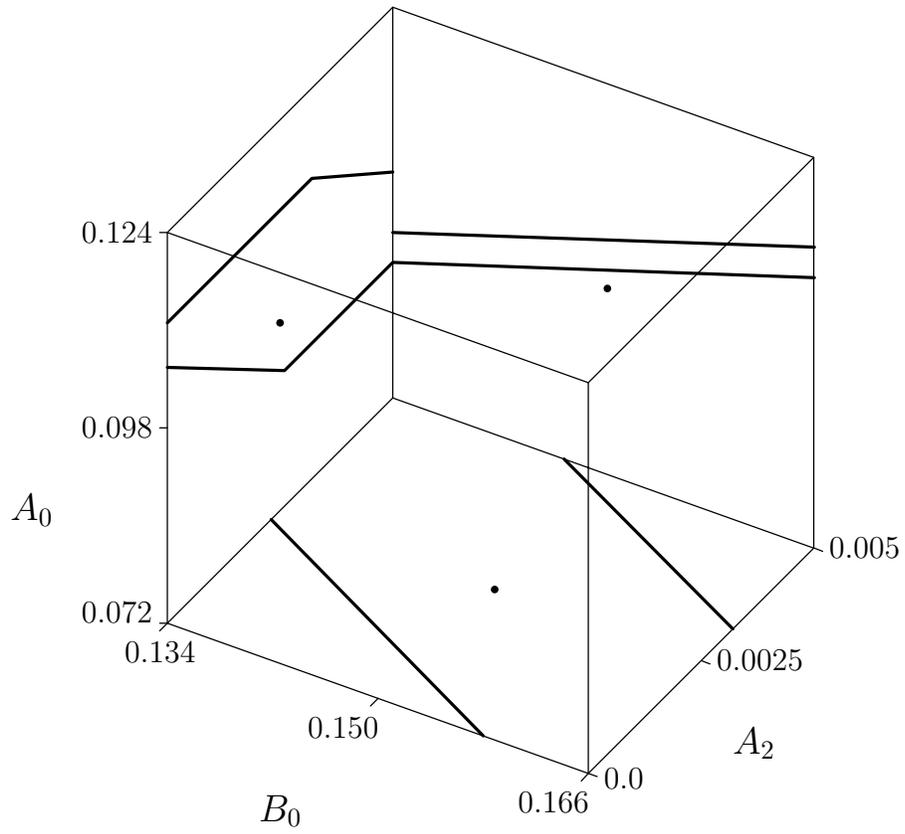
\noindent {\bf Fig. 3.} The projections of $\overline{\Delta}$
and the central experimental values onto the $(A_0,A_2)$-,
$(A_2,B_0)$- and $(A_0,B_0)$- planes.  The dots represent the
central experimental values obtained from (3.4).


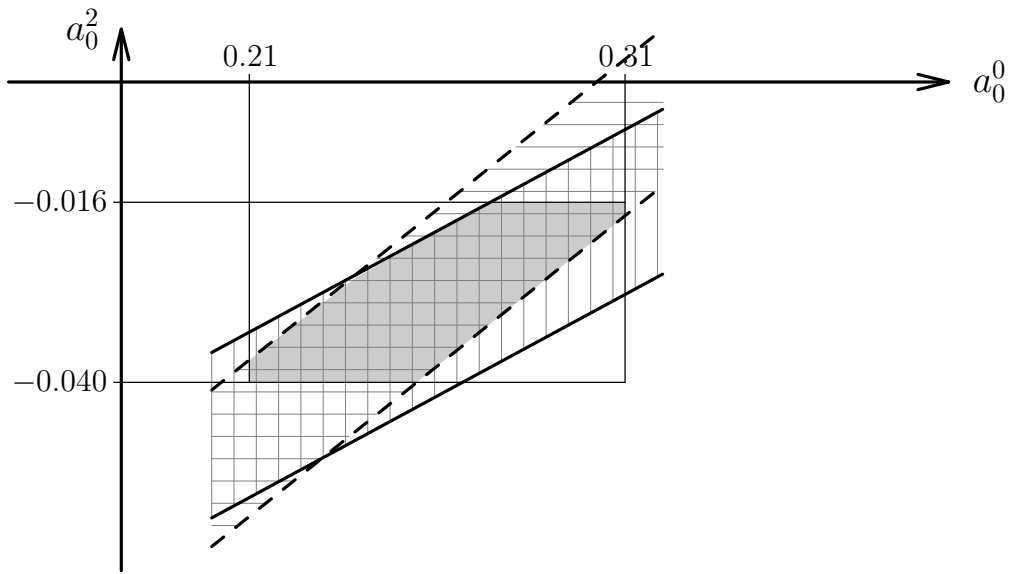
\noindent {\bf Fig. 4.} Constraints on the S- waves scattering
lengths.  The experimental data define a rectangle of allowed
values, the ``universal'' strip~\cite{Nagels} is
bounded by the dashed lines $a^2_0=0.4 a^0_0-0.131\pm 0.010$ and
our band is limited by the full lines $a^2_0=0.27 a^0_0-0.101 \pm 0.011$.
The constraints confine $(a^0_0, a^2_0)$ to the shaded area.


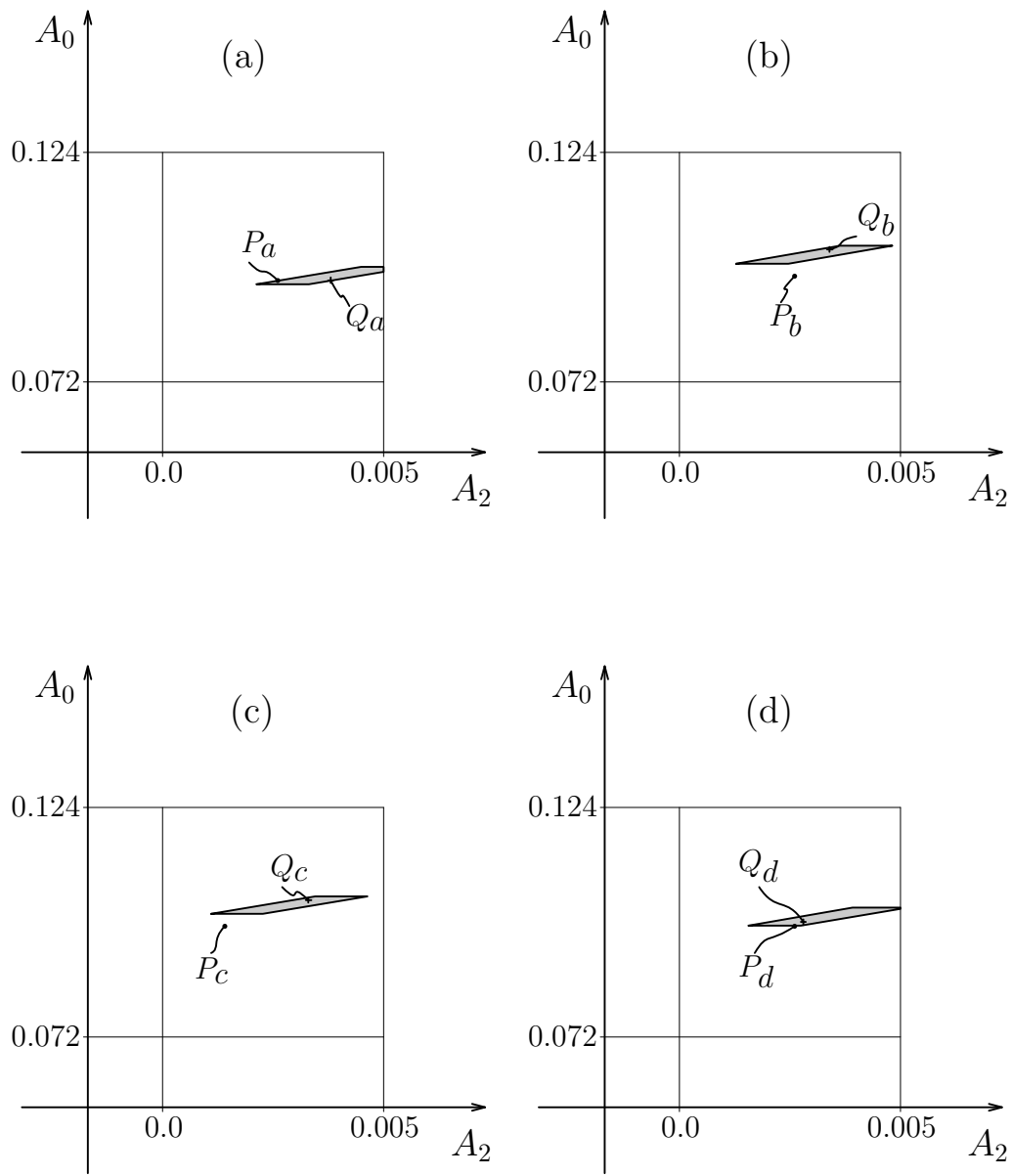
\noindent {\bf Fig. 5.} Sections of $\Delta$ and
$\overline{\Delta}$ at five fixed values of $B_0$ obtained
from CHPT, {\bf (a)} one loop SCHPT, $B_0=0.140$, {\bf (b)-(c)}
one loop GCHPT, $B_0=0.149,\ 0.151$, {\bf (d)} two loop GCHPT,
$B_0=0.146$.  The significance of the points $Q$ and $P$ is explained
in the text.

\newpage

\begin{footnotesize}
$$
\begin{array}{||c|c|c|c|c|c|c|c|c|c|c||}\hline
\alpha & C_{\alpha} & R_{\alpha 0} & R_{\alpha 1} & R_{\alpha 2} &
S_{\alpha 0} & S_{\alpha 1} & S_{\alpha 2} &
T_{\alpha 0} & T_{\alpha 1} & T_{\alpha 2}   \\ \hline
1 & 0.00 & 0.07 & -0.06 & 0.18 & 0.08 & 1.19 & 0.03 & 0.68 &
10.28 & -0.01 \\
2 & 0.02 & 0.63 & 0.33 &  1.93 & -0.41 & -0.89 & -0.59 & 4.03 &
-16.15 & -0.16 \\
3 & 0.00 & 0.07 & -0.10 & 0.18 & 0.08 & 0.31 & 0.03 & 0.68 &
0.10 & -0.01 \\  \hline
\end{array}
$$
\bigskip
\vskip 2cm
$$
\begin{array}{||c|c|c|c|c|c|c||}\hline
\alpha &
U_{\alpha 0} & U_{\alpha 1} & U_{\alpha 2} &
V_{\alpha 0} & V_{\alpha 1} & V_{\alpha 2}   \\ \hline
1 & 0.02 & -12.42 & -0.11 & 0.09 & 10.11 & -0.69 \\
2 &-0.16 & -240.3 & 0.75 & -2.49 & 1.24 & 0.58 \\
3 & 0.02 & -7.67 & -0.11 & 0.09 & -1.82 & -0.69 \\
\hline
\end{array}
$$
\bigskip
\begin{center}
{\Large{\rm Table\ 1}}
\end{center}
\newpage
\begin{center}
\begin{tabular}{||c|c| c|c |c|c |c||}\hline
 &\multicolumn{2}{c|}{SCHPT one loop} & \multicolumn{4}{c||} {GCHPT one loop}
 \\ \cline{4-7}
& \multicolumn{2}{c|} {(a)} & \multicolumn{2}{c|} {(b)} &
 \multicolumn{2}{c||} {(c)} \\ \cline{2-7}
& & {Sum rules} & & {Sum rules} & & {Sum rules} \\ \hline
$a^0_0$ & 0.20 & & 0.27 & & 0.28 & \\
$b^0_0$ & 0.25 & & 0.26 & & 0.28 & \\
$a^1_1$ & 0.037 & & 0.039 & & 0.039 & \\
$A_0$ & 0.095 & 0.095 & 0.096 & 0.102 & 0.097 & 0.099 \\
$B_0$ & 0.140 & & 0.149 & & 0.151 & \\
$A_2$ & 0.0026 & 0.0038 & 0.0024 & 0.0036 & 0.0014 & 0.0028 \\
$b^1_1$ & 0.0044 & 0.0056 & 0.0048 & 0.0054 & 0.0028 & 0.0034 \\ \hline
\end{tabular}

\bigskip
\vskip 1cm

\begin{tabular}{||c|c|c||}\hline
& \multicolumn{2}{c||} {GCHPT two loops} \\
& \multicolumn{2}{c||} {(d)} \\ \cline{2-3}
& & {Sum rules} \\ \hline
$a^0_0$ & 0.26 & \\
$b^0_0$ & 0.25 & \\
$a^1_1$ & 0.037 & \\
$A_0$ & 0.097 & 0.098 \\
$B_0$ & 0.146 & \\
$A_2$ & 0.0026 & 0.0028 \\
$b^1_1$ & 0.0054 & 0.0056 \\ \hline
\end{tabular}

\bigskip
\vskip 1cm
{\Large Table 2}
\end{center}

\end{footnotesize}


\newpage
\parskip=4mm
\parindent=0cm

\centertexdraw{
\drawdim mm
\linewd 0.4
\lpatt(2 2)\move(27 48)\lvec(65 27)\lpatt()\lvec(56 15.4)\lvec(0 13.3)
\lvec(27 48)\lfill f:0.3
\move(0 21)\lvec(56 18)\lvec(86 48)\lvec(30 51)\lvec(0 21)
\lfill f:0.8
\lpatt(2 2)\move(27 48)\lvec(65 27)\lpatt()\lvec(86 54)\lvec(30 51.8)
\lvec(27 48)\lfill f:0.3
\linewd 0.2
\move(0 0)
\rlvec(0 52)
\rlvec(56 -20)\rlvec(0 -52)\rlvec(-56 20)
\rlvec(30 30)\rlvec(0 52)\rlvec(56 -20)\rlvec(0 -52)\rlvec(-56 20)
\move(0 52)\rlvec(30 30)\move(56 32)\rlvec(30 30)\rmove(0 -52)\rlvec(-30 -30)
\move(-1 0)\lvec(0 0)
\move(-1 26)\lvec(0 26)
\move(-1 52)\lvec(0 52)
\move(-1 -1)\lvec(0 0)
\move(27 -11)\lvec(28 -10)
\move(55 -21)\lvec(56 -20)
\move(57.12 -20.4)\lvec(56 -20)
\move(72.12 -5.4)\lvec(71 -5)
\move(87.12 9.6)\lvec(86 10)
\textref h:R v:C
\htext(-2 26){0.098}
\htext(-2 52){0.124}
\textref h:R v:B
\htext(-2 0){0.072}
\textref h:C v:T
\htext(-1 -2.5){0.134}
\textref h:R v:T
\htext(28 -12.5){0.150}
\htext(56 -22.5){0.166}
\textref h:L v:B
\htext(58 -22){0.0}
\textref h:L v:C
\htext(73 -5){0.0025}
\htext(88 10){0.005}
\textref h:C v:C
\htext(-18 15){\large $A_0$}
\htext(15 -25){\large $B_0$}
\htext(78 -16){\large $A_2$}
\htext(35 -90){\large Fig. 1}
}

\vspace{10cm}
\centertexdraw{
\drawdim mm

\linewd 0.1
\move(0 0)
\lvec(51 -10)\lvec(51 80)\lvec(0 70)\lvec(0 0)
\lpatt(2 2)\lvec(32 4)\lvec(82 -3)
\lpatt()\lvec(82 73)
\lpatt(2 2)\lvec(32 66)\lvec(32 4)
\lpatt()\move(51 -10)\lvec(82 -3)
\move(51 80)\lvec(82 73)
\move(0 70)\lpatt(2 2)\lvec(32 66)

\lpatt()
\linewd 0.2
\move(16 26)\fcir f:0.5 r:0.5   \lvec(22.5 26.5)\fcir f:0.5 r:0.5
  \lvec(32 30.6)\lpatt(.5 .8)
\lvec(32 32)\lpatt(1 1)
\lvec(28 32)\fcir f:0.5 r:0.5   \lvec(16 26)
\lpatt()\lvec(43.1 47.2)\fcir f:0 r:0.5   \linewd 0.4
\lvec(51 53)\lvec(51 52)\lvec(54 51.5)\fcir f:0 r:0.5
  \lvec(66.2 57.2)\fcir f:0 r:0.5
\lvec(59.1 58)\fcir f:0 r:0.5   \lvec(51 53)
\move(51 52)\lvec(43.1 47.2)\linewd 0.2
\move(66.2 57.2)\lvec(40 35)\fcir f:0.5 r:0.5   \lpatt(1 1)\lvec(32 32)
\move(32 30.6)\lpatt()\lvec(40 35)
\move(28 32)\lpatt(1 1)\lvec(59.1 58)
\move(22.5 26.5)\lpatt()\lvec(54 51.5)
\linewd 0.4
\move(43.1 47.2)
\lvec(51 53)\lvec(51 52)\lvec(54 51.5)\lvec(66.2 57.2)
\lvec(59.1 58)\lvec(51 53)
\move(51 52)\lvec(43.1 47.2)
\move(43.1 47.2)\fcir f:0 r:0.5
\move(54 51.5)\fcir f:0 r:0.5

\move(16 26)\fcir f:0 r:0.5
\move(22.5 26.5)\fcir f:0 r:0.5
\move(28 32)\fcir f:0 r:0.5
\move(40 35)\fcir f:0 r:0.5
\linewd 0.2

\move(-1 0)\lvec(0 0)
\move(-1 35)\lvec(0 35)
\move(-1 70)\lvec(0 70)
\move(-1 -1)\lvec(0 0)
\move(24.5 -6)\lvec(25.5 -5)
\move(50 -11)\lvec(51 -10)
\move(52.12 -10.4)\lvec(51 -10)
\move(67.62 -6.8)\lvec(66.5 -6.4)
\move(83.12 -3.4)\lvec(82 -3)
\textref h:R v:C
\htext(-2 35){0.098}
\htext(-2 70){0.124}
\textref h:R v:B
\htext(-2 0){0.072}
\textref h:C v:T
\htext(-1 -2.5){0.134}
\textref h:R v:T
\htext(25.5 -7.5){0.150}
\htext(50.5 -12.5){0.166}
\textref h:L v:B
\htext(53 -13){0.0}
\textref h:L v:C
\htext(68.5 -8.4){0.0025}
\htext(84 -5){0.005}
\textref h:C v:C
\htext(-18 40){\large $A_0$}
\htext(15 -15){\large $B_0$}
\htext(78 -16){\large $A_2$}
\htext(35 -70){\large Fig. 2}
}

\vspace{10cm}
\centertexdraw{
\drawdim mm
\linewd 0.4
\move(0 40)\lvec(19.2 59.2)\lvec(30 60)             
\move(0 34)\lvec(15.6 33.6)\lvec(30 48)\lvec(86 46) 
\move(75.2 -0.8)\lvec(52.75 21.875)                 
\move(42 -15)\lvec(13.8 13.8)                       
\move(30 52)\lvec(86 50)                            
\linewd 0.2
\move(15 40)\fcir f:0 r:0.5
\move(43.5 4.5)\fcir f:0 r:0.5
\move(58.5 44.5)\fcir f:0 r:0.5

\move(0 0)
\rlvec(0 52)
\rlvec(56 -20)\rlvec(0 -52)\rlvec(-56 20)
\rlvec(30 30)\rlvec(0 52)\rlvec(56 -20)\rlvec(0 -52)\rlvec(-56 20)
\move(0 52)\rlvec(30 30)\move(56 32)\rlvec(30 30)\rmove(0 -52)\rlvec(-30 -30)
\move(-1 0)\lvec(0 0)
\move(-1 26)\lvec(0 26)
\move(-1 52)\lvec(0 52)
\move(-1 -1)\lvec(0 0)
\move(27 -11)\lvec(28 -10)
\move(55 -21)\lvec(56 -20)
\move(57.12 -20.4)\lvec(56 -20)
\move(72.12 -5.4)\lvec(71 -5)
\move(87.12 9.6)\lvec(86 10)
\textref h:R v:C
\htext(-2 26){0.098}
\htext(-2 52){0.124}
\textref h:R v:B
\htext(-2 0){0.072}
\textref h:C v:T
\htext(-1 -2.5){0.134}
\textref h:R v:T
\htext(28 -12.5){0.150}
\htext(56 -22.5){0.166}
\textref h:L v:B
\htext(58 -22){0.0}
\textref h:L v:C
\htext(73 -5){0.0025}
\htext(88 10){0.005}
\textref h:C v:C
\htext(-18 15){\large $A_0$}
\htext(15 -25){\large $B_0$}
\htext(78 -16){\large $A_2$}
\htext(35 -90){\large Fig. 3}
}

\vspace{10cm}
\centertexdraw{
\drawdim mm
\arrowheadtype t:V
\move(17 -37)\lvec(17 -40)\lvec(39.25 -40)\lvec(67 -18)
\lvec(67 -16)\lvec(49.04 -16)\lvec(31.23 -26)\ifill f:0.8
\setgray 0.5
\linewd 0.1
\move(12 -62)\rlvec(60 0)
\rmove(-60 3)\rlvec(60 0) \rmove(-60 3)\rlvec(60 0)
\rmove(-60 3)\rlvec(60 0) \rmove(-60 3)\rlvec(60 0)
\rmove(-60 3)\rlvec(60 0) \rmove(-60 3)\rlvec(60 0)
\rmove(-60 3)\rlvec(60 0) \rmove(-60 3)\rlvec(60 0)
\rmove(-60 3)\rlvec(60 0) \rmove(-60 3)\rlvec(60 0)
\rmove(-60 3)\rlvec(60 0) \rmove(-60 3)\rlvec(60 0)
\rmove(-60 3)\rlvec(60 0) \rmove(-60 3)\rlvec(60 0)
\rmove(-60 3)\rlvec(60 0) \rmove(-60 3)\rlvec(60 0)
\rmove(-60 3)\rlvec(60 0) \rmove(-60 3)\rlvec(60 0)
\rmove(-60 3)\rlvec(60 0) \rmove(-60 3)\rlvec(60 0)
\move(12 -63)\lvec(12 -61.8)\lvec(72 -13.8)\lvec(73 -13.8)\lvec(73 -63)
\ifill f:1
\move(12 -41)\lvec(72 7)\lvec(72 1)\lvec(11 1)\lvec(11 -41)
\ifill f:1
\move(12 -36)\rlvec(0 -22)
\rmove(3 23.62)\rlvec(0 -22) \rmove(3 23.62)\rlvec(0 -22)
\rmove(3 23.62)\rlvec(0 -22) \rmove(3 23.62)\rlvec(0 -22)
\rmove(3 23.62)\rlvec(0 -22) \rmove(3 23.62)\rlvec(0 -22)
\rmove(3 23.62)\rlvec(0 -22) \rmove(3 23.62)\rlvec(0 -22)
\rmove(3 23.62)\rlvec(0 -22) \rmove(3 23.62)\rlvec(0 -22)
\rmove(3 23.62)\rlvec(0 -22) \rmove(3 23.62)\rlvec(0 -22)
\rmove(3 23.62)\rlvec(0 -22) \rmove(3 23.62)\rlvec(0 -22)
\rmove(3 23.62)\rlvec(0 -22) \rmove(3 23.62)\rlvec(0 -22)
\rmove(3 23.62)\rlvec(0 -22) \rmove(3 23.62)\rlvec(0 -22)
\rmove(3 23.62)\rlvec(0 -22) \rmove(3 23.62)\rlvec(0 -22)
\setgray 0
\linewd 0.4
\move(12 -36)\lvec(72 -3.6)
\move(12 -58)\lvec(72 -25.6)
\lpatt(2 2)\move(12 -41)\lvec(72 7)
\move(12 -61.8)\lvec(72 -13.8)\lpatt()
\linewd 0.4
\move(-15 0)\avec(110 0)
\move(0 -65)\avec(0 7)
\linewd 0.2
\move(-1 -16)\lvec(67 -16) \move(17 1)\lvec(17 -40)
\move(-1 -40)\lvec(67 -40) \move(67 1)\lvec(67 -40)
\textref h:R v:C
\htext(-2 -16){$-0.016$}
\htext(-2 -40){$-0.040$}
\htext(-3 7){\large $a_0^2$}
\textref h:L v:C
\htext(113 0){\large $a_0^0$}
\textref h:C v:B
\htext(17 2){$0.21$}
\htext(67 2){$0.31$}
\htext(40 -90){\large Fig. 4}
}

\twocolumn
\btexdraw
\drawdim mm \setunitscale 0.6
\arrowheadtype t:V
\arrowheadsize l:3 w:1.5

\move(38.3 38.06)\lvec(50.1 38.06)\lvec(67 40.92)\lvec(67 42.06)
\lvec(61.9 42.06)\lvec(38.3 38.06)\ifill f:0.8
\setgray 0
\move(38.3 38.06)\lvec(50.1 38.06)\lvec(67 40.92)\lvec(67 42.06)
\lvec(61.9 42.06)\lvec(38.3 38.06)
\move(43 39)\fcir f:0 r:0.5
\clvec(39 43)(40.5 39)(36.5 43)
\htext(34.5 44){$P_{\displaystyle a}$}

\move(55 38.4)\lvec(55 39.6)
\move(54.4 39)\lvec(55.6 39)
\move(55 39)\clvec(59.2 31.1)(56.2 39.1)(59.2 33.1)
\htext(58.2 27.1){$Q_{\displaystyle a}$}
\linewd 0.4
\move(-15 0)\avec(90 0)
\move(0 -15)\avec(0 100)
\linewd 0.2
\move(-1 16)\lvec(67 16) \move(17 -1)\lvec(17 68)
\move(-1 68)\lvec(67 68) \move(67 -1)\lvec(67 68)
\textref h:R v:C
\htext(-2 16){$0.072$}
\htext(-2 68){$0.124$}
\htext(-3 95){\large $A_0$}
\textref h:L v:C
\htext(82 -9){\large $A_2$}
\textref h:C v:B
\htext(17 -7){$0.0$}
\htext(67 -7){$0.005$}
\htext(35 85){\large (a)}
\etexdraw

\vspace{2cm}

\btexdraw
\drawdim mm \setunitscale 0.6
\arrowheadtype t:V
\arrowheadsize l:3 w:1.5

\move(27.9 43.88)\lvec(39.7 43.88)\lvec(63.3 47.88)\lvec(51.5 47.88)
\lvec(27.9 43.88)\ifill f:0.8
\setgray 0
\move(27.9 43.88)\lvec(39.7 43.88)\lvec(63.3 47.88)\lvec(51.5 47.88)
\lvec(27.9 43.88)
\move(31 41)\fcir f:0 r:0.5
\clvec(28 38)(30 36)(28 35)
\htext(24 28){$P_{\displaystyle c}$}

\move(50 46.4)\lvec(50 47.6)
\move(49.4 47)\lvec(50.6 47)
\move(50 47)\clvec(45 52)(49 45)(44 50)
\htext(41 51){$Q_{\displaystyle c}$}

\linewd 0.4
\move(-15 0)\avec(90 0)
\move(0 -15)\avec(0 100)
\linewd 0.2
\move(-1 16)\lvec(67 16) \move(17 -1)\lvec(17 68)
\move(-1 68)\lvec(67 68) \move(67 -1)\lvec(67 68)
\textref h:R v:C
\htext(-2 16){$0.072$}
\htext(-2 68){$0.124$}
\htext(-3 95){\large $A_0$}
\textref h:L v:C
\htext(82 -9){\large $A_2$}
\textref h:C v:B
\htext(17 -7){$0.0$}
\htext(67 -7){$0.005$}
\htext(37 85){\large (c)}
\etexdraw

\pagebreak

\btexdraw
\drawdim mm \setunitscale 0.6
\arrowheadtype t:V
\arrowheadsize l:3 w:1.5

\move(29.8 42.82)\lvec(41.6 42.82)\lvec(65.2 46.82)\lvec(53.4 46.82)
\lvec(29.8 42.82)\ifill f:0.8
\setgray 0
\move(29.8 42.82)\lvec(41.6 42.82)\lvec(65.2 46.82)\lvec(53.4 46.82)
\lvec(29.8 42.82)
\move(43 40)\fcir f:0 r:0.5
\clvec(37.5 34)(44 40)(40 34)
\htext(37 26){$P_{\displaystyle b}$}

\move(51 45.4)\lvec(51 46.6)
\move(50.4 46)\lvec(51.6 46)
\move(51 46)\clvec(55 47)(52 48)(57 49)
\htext(57 49){$Q_{\displaystyle b}$}

\linewd 0.4
\move(-15 0)\avec(90 0)
\move(0 -15)\avec(0 100)
\linewd 0.2
\move(-1 16)\lvec(67 16) \move(17 -1)\lvec(17 68)
\move(-1 68)\lvec(67 68) \move(67 -1)\lvec(67 68)
\textref h:R v:C
\htext(-2 16){$0.072$}
\htext(-2 68){$0.124$}
\htext(-3 95){\large $A_0$}
\textref h:L v:C
\htext(82 -9){\large $A_2$}
\textref h:C v:B
\htext(17 -7){$0.0$}
\htext(67 -7){$0.005$}
\htext(37 85){\large (b)}
\etexdraw

\vspace{2cm}

\btexdraw
\drawdim mm \setunitscale 0.6
\arrowheadtype t:V
\arrowheadsize l:3 w:1.5

\move(32.6 41.23)\lvec(44.4 41.23)\lvec(67 45.05)\lvec(67 45.23)\lvec(56.2
45.23)
\lvec(32.6 41.23)\ifill f:0.8
\setgray 0
\move(32.6 41.23)\lvec(44.4 41.23)\lvec(67 45.05)\lvec(67 45.23)
\lvec(56.2 45.23)
\lvec(32.6 41.23)

\move(43 41)\fcir f:0 r:0.5
\clvec(37 37)(37 40)(34 35)
\htext(30 27){$P_{\displaystyle d}$}

\move(45 41.4)\lvec(45 42.6)
\move(44.4 42)\lvec(45.6 42)
\move(45 42)\clvec(41 48)(39 44)(35 50)
\htext(30 51){$Q_{\displaystyle d}$}

\linewd 0.4
\move(-15 0)\avec(90 0)
\move(0 -15)\avec(0 100)
\linewd 0.2
\move(-1 16)\lvec(67 16) \move(17 -1)\lvec(17 68)
\move(-1 68)\lvec(67 68) \move(67 -1)\lvec(67 68)
\textref h:R v:C
\htext(-2 16){$0.072$}
\htext(-2 68){$0.124$}
\htext(-3 95){\large $A_0$}
\textref h:L v:C
\htext(82 -9){\large $A_2$}
\textref h:C v:B
\htext(17 -7){$0.0$}
\htext(67 -7){$0.005$}
\htext(37 85){\large (d)}
\etexdraw

\vspace{1cm}
\hspace{-0.9cm}
{\large Fig. 5}

\end{document}